\documentclass[11pt]{article}

\usepackage{graphicx}
\usepackage{txfonts}
\usepackage{epstopdf}
\usepackage{subcaption}
\usepackage{color,colortbl}
\usepackage{array}
\usepackage{url}
\usepackage[numbers]{natbib}
\usepackage{hyperref}
\usepackage{authblk}

\newcommand{\CII}{[C{\scriptsize\,II}]}
\newcommand{\TCII}{[$^{13}$C{\scriptsize\,II}]}
\newcommand{\CI}{[C{\scriptsize\,I}]}
\newcommand{\OI}{[O{\scriptsize\,I}]}

\newcommand{\HI}{H{\scriptsize\,I}}

\makeatletter
\renewcommand{\@seccntformat}[1]{}
\makeatother

\begin{document}

\title{Ionized carbon as a tracer of the assembly of interstellar clouds}


\author[1]{Nicola Schneider} 
\author[2]{Lars Bonne} 
\author[3]{Sylvain Bontemps} 
\author[1]{Slawa Kabanovic} 
\author[1]{RobertSimon} 
\author[1]{Volker Ossenkopf-Okada} 
\author[1]{Christof Buchbender} 
\author[1]{J\"urgen Stutzki} 
\author[1]{Marc Mertens} 
\author[4]{Oliver Ricken} 
\author[3]{Timea Csengeri} 
\author[5,6]{Alexander G.G.M. Tielens} 


\affil[1]{I. Physikalisches Institut, Universit\"at zu K\"oln, Z\"ulpicher Str. 77, Cologne, Germany}
\affil[2]{SOFIA Science Center, NASA Ames Research Center, Moffet Field, CA, USA}
\affil[3]{Laboratoire d'Astrophysique de Bordeaux, Universit\'e de Bordeaux, Pessac, France}
\affil[4]{Max-Planck Institut f\"ur Radioastronomie, Bonn, Germany}
\affil[5]{Department of Astronomy, University of Maryland, College Park, USA}
\affil[6]{Leiden Observatory, Leiden University, Leiden, The Netherlands}

\maketitle

\abstract{Molecular hydrogen clouds are a key component of the
  interstellar medium because they are the birthplaces for stars.
  They are embedded in atomic gas that pervades the interstellar
  space.  However, the details of how molecular clouds assemble from
  and interact with the atomic gas are still largely unknown.  As a
  result of new observations of the 158~$\mu$m line of ionized carbon
  \CII\ in the Cygnus region within the FEEDBACK program on SOFIA
  (Stratospheric Observatory for Infrared Astronomy), we present 
  compelling evidence that \CII\ unveils dynamic 
  interactions between cloud ensembles.  This process is neither a
  head-on collision of fully molecular clouds nor a gentle merging of
  only atomic clouds. Moreover, we demonstrate that the dense
  molecular clouds associated with the DR21 and W75N star-forming
  regions and a cloud at higher velocity are embedded in atomic gas
  and all components interact over a large range of velocities
  ($\sim$20 km s$^{-1}$). The atomic gas has a density of $\sim$100
  cm$^{-3}$ and a temperature of $\sim$100 K.  We conclude that the
  \CII\ 158 $\mu$m line is an excellent tracer to witness the
  processes involved in cloud interactions and anticipate further
  detections of this phenomenon in other regions.}

\section{}\label{intro}

Molecular clouds are a crucial component of the interstellar medium
(ISM) of galaxies as they are the birth sites of stars and planetary
systems. However, the processes by which these clouds are assembled
from the large atomic hydrogen (\HI) reservoir in galaxies is still
not well understood. Some models are based on a subtle equilibrium
between gravity, turbulence and magnetic fields
\citep[e.g.][]{Krumholz2005}. An external increase of pressure or
turbulence due to stellar feedback or spiral arm density waves then
randomly triggers a quasi-static, slow build-up of density, leading to
the formation of pockets of gas of molecular hydrogen (H$_2$).  Other
models \citep[e.g.][]{Hartmann2001} propose that cloud formation is
more dynamic and driven by large scale motions in the galaxy, but
still closely linked to the local transition from warm (T$\sim$5000
K), tenuous, mostly atomic gas to dense, cooler (T$\lesssim$100 K),
partly molecular gas.  In this simple two-phase model of the ISM, only
the warm and cold neutral medium (WNM and CNM, respectively) are
thermally stable. Gas at intermediate temperatures is not in
equilibrium and, depending on its density, will either cool down and
become denser and fully molecular or heat up to join the WNM. In
addition, stellar feedback effects such as radiation, winds, and
supernova explosions generate turbulence and complicate the
picture. It is thus challenging to find the right observational
tracers for both, the dynamic interaction between gas flows and the
thermal and chemical transitions between WNM and CNM.

In simulations, dynamic cloud formation scenarios are idealized by low
velocity ($\lesssim$10 km s$^{-1}$) converging flows
\citep[e.g.][]{Koyama2000,Vazquez2006,Inoue2009,Clark2019} which 
convert diffuse \HI\ gas into dense H$_2$ gas. A recent study
\citep{Dobbs2020} showed that only flows with hydrogen densities
$\simeq$100 cm$^{-3}$ that collide with velocities $\simeq$20 km
s$^{-1}$ manage to build up massive structures in which stellar
proto-clusters can form.  In models with even higher density, the gas
flows are already molecular before they collide and are then referred
to "cloud-cloud collisions" (CCCs)
\citep{Haworth2015,Bisbas2017,Fukui2021}.  Observations with
velocities $\gtrsim$20 km $^{-1}$ are reported in 
\citep{Fukui2014,Torii2015}.  However, these different scenarios
result in contrasting observational predictions.  Colliding \HI\ flow
models \citep{Clark2019} anticipate many velocity components in the
lines of ionized carbon (\CII) and much less in the rotational
transitions of carbon monoxide (CO).  \CII\ emission has its origin in
the atomic gas and from non-thermal contributions of multiple
molecular clump surfaces at different velocities along the
line-of-sight, while CO only arises from the molecular component.
Cloud-cloud collision simulations \citep{Haworth2015} produce two main molecular
velocity components visible in CO, with a bridge of emission in
velocity space in between the two components. \CII\ emission stems
mostly from the envelope of the molecular cloud and the surrounding
ambient ISM gas that does not participate in the collision
\citep{Bisbas2017}.

How can these differing views be confronted with observations?  The
21cm line of \HI\ can be observed in emission and absorption, but it
mostly fills the interstellar space so that velocity information is
highly blurred. CO lines are used as a proxy for H$_2$ in dense, fully
molecular clouds. However, because H$_2$ self-shields more effectively
from ultraviolet (UV) photodissociation than CO, there is a gas component that is
mostly dark in CO, but bright in H$_2$
\citep{Wolfire2010}. Fortunately, the \CII\ fine-structure line at 158
$\mu$m is perfectly suited to determine the physical conditions in
atomic and CO-dark molecular gas
\citep{Pineda2013,Beuther2014,Franeck2018} and is hence an excellent
observational tracer for molecular cloud formation frameworks.

%
\begin{figure*}[htp]
\begin{center} 
\includegraphics [width=12cm, angle={0}]{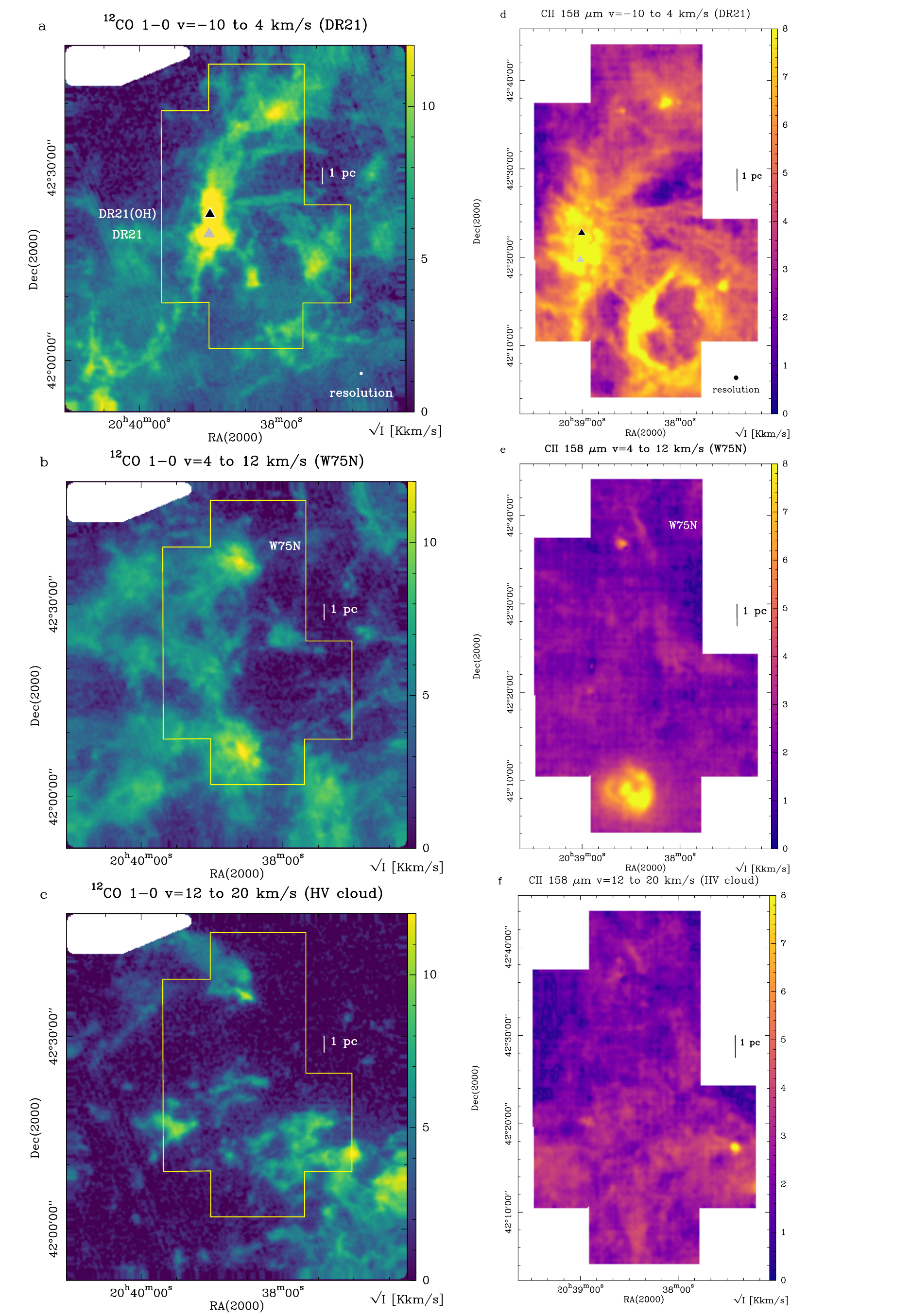}
\caption{Velocity integrated CO and \CII\ emission in the DR21, W75N
  and high-velocity range. {\bf a-c} $^{12}$CO 1$\to$0 maps obtained
  with the Nobeyama telescope \citep{Yamagishi2018} in the three main
  velocity ranges of the Cygnus X region: $-$10-4 km s$^{-1}$ (a), $-$4-12 km s$^{-1}$ (b), and 
  12-20 km s$^{-1}$ (c).  The embedded massive 
  star-forming sites DR21, DR21(OH) and W75N are indicated. The yellow
  polygon outlines the area mapped in \CII\ that is shown in panels
  {\bf d-f}. The color wedges give the CO and \CII\ intensities in
  square root values. The resolution of the maps (30$''$) is indicated
  in panels a and d. RA, right acension; Dec, Declination.}
\label{fig1}
\end{center} 
\end{figure*}

We here present observations in the \CII\ 158 $\mu$m and CO 1$\to$0
spectral lines of Cygnus X, a region with a coherent network of
molecular clouds \citep{Schneider2006}, extending over $\sim$100-200
pc. It includes the massive star-forming regions DR21 and W75N at a
distance of 1.5$\pm$0.1 and 1.3$\pm$0.1 kpc, respectively,
determined with maser parallax measurements \citep{Rygl2012}, and the
rich Cyg OB2 association with 169 OB stars \citep{Wright2015}. Massive
star formation occurs in parts of the clouds, for example, the young stellar
outflow source DR21, the cluster-forming site DR21(OH), and the
cluster of early type B stars in W75N \citep{Reipurth2008}. Cygnus X
is not exceptional in terms of mass \citep{Quang2016}, but for its
stellar content. Using $^{12}$CO 1$\to$0 observations, it was
suggested \citep{Dickel1978,Dobashi2019} that the DR21 molecular
cloud, that has a systemic velocity v = $-$3 km s$^{-1}$ with respect
to the local standard of rest, collides with the W75N molecular cloud
(v = 9 km s$^{-1}$).  The \CII\ observations reported here indicate a
different scenario with an interaction between atomic and molecular
gas over a large range of velocities ($\sim$20 km s$^{-1}$).

As part of the SOFIA (Stratospheric Observatory for Infrared
Astronomy) FEEDBACK legacy program, the Cygnus X region was observed
with the heterodyne array receiver upGREAT (Methods)
\citep{Risacher2018} in the 158 $\mu$m line of \CII\ at an angular
resolution of 14.1$''$, corresponding to 0.1 pc at a distance of 1.4
kpc. This map of $\sim$950 arcmin$^2$ (158 pc$^2$) size is, 
together with the SOFIA map of Orion A \citep{Pabst2019}, a very large, 
high angular resolution \CII\ map showing a very massive
star-forming region that extends well into the outskirts of the
molecular clouds. It provides data at sub-km s$^{-1}$ spectral
resolution so that, for the first time, the dynamic assemblage of
molecular clouds can be traced in detail.

We smoothed the \CII\ map to an angular resolution of 30$''$ and
re-gridded to 10$''$.  The data were resampled to a velocity
resolution of 0.5 km s$^{-1}$ and have a mean noise temperature of 0.3
K per channel (Methods and Extended Data Fig.~\ref{noise}). We also
use $^{12}$CO 1$\to$0 data from the Nobeyama Cygnus survey
\citep{Yamagishi2018}, smoothed to 30$''$ resolution on a 10$''$ grid
and a mean noise per 0.5 km s$^{-1}$ wide channel of 0.6~K, and
\HI\ data from the Canadian Galactic Plane survey
\citep{Taylor2003}. The \HI\ data have an angular resolution of 1$'$
and a root mean square (r.m.s.) noise of 3 K in a 0.82 km s$^{-1}$ channel.

\section{Results}\label{results}

\subsection{Distribution of CO and \CII\ emission in Cygnus X} \label{large}

\begin{figure*}[ht]
\begin{center} 
\includegraphics [width=13cm, angle={0}]{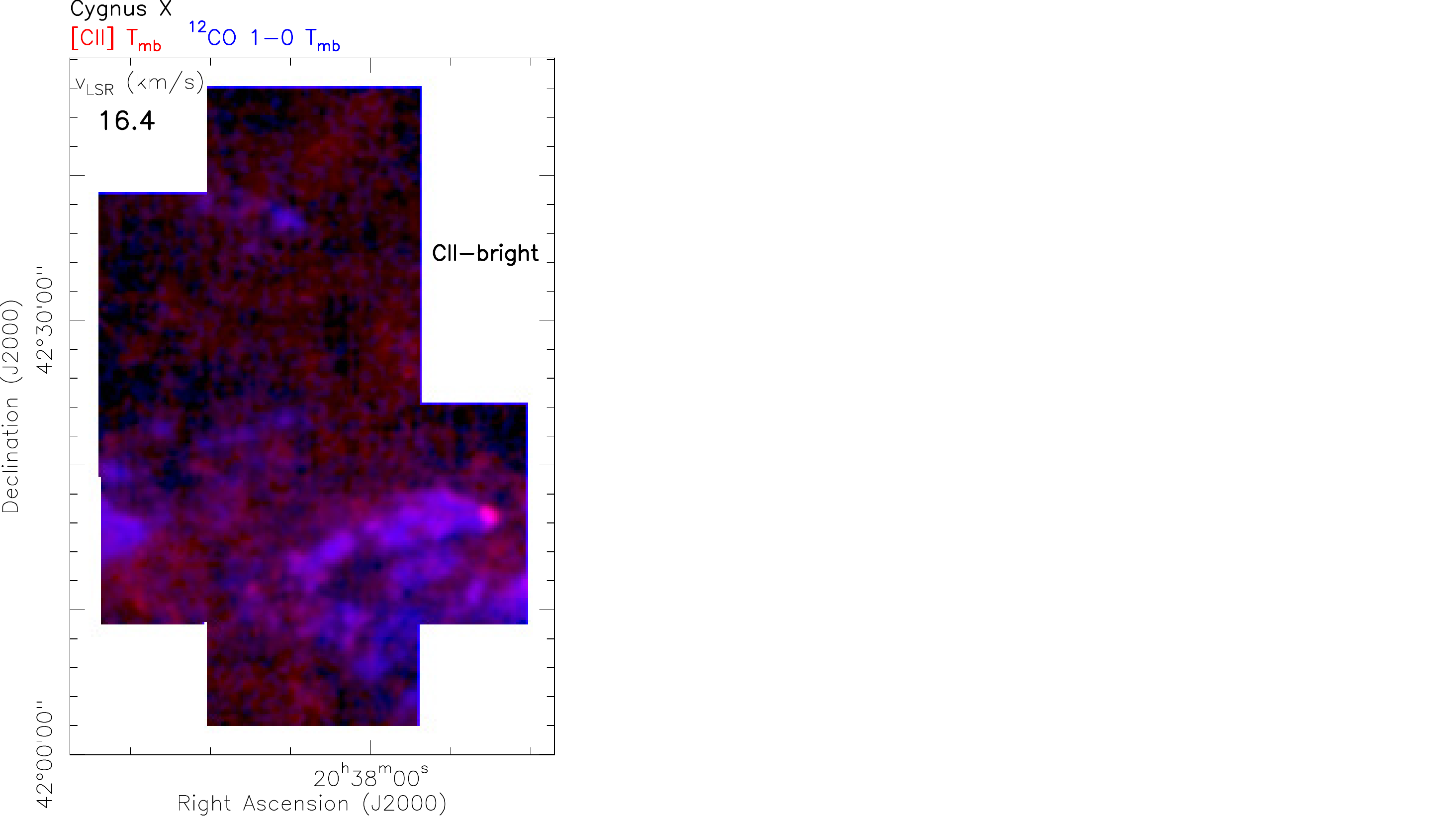}
\caption{Movie-snapshot of \CII\ and CO emission. The image displays a
  single velocity channel (16.4 km s$^{-1}$) of \CII\ emission in red
  (0 to 16 K km s$^{-1}$) and CO emission in blue (0 to 25 K km
  s$^{-1}$), respectively, plotted as square-root values. Areas that
  are dark in CO (no or little emission) are bright in \CII\ and
  reveal the large spatial extent of \CII\ emission. The full movie
  with all velocity channels is found at
  \url{https://astro.uni-koeln.de/stutzki/research/feedback/animations}.}\label{fig2movie}
\end{center} 
\end{figure*}

%
Figure~\ref{fig1} shows the $^{12}$CO 1$\to$0 and \CII\ line
integrated emission distribution in the three major velocity ranges in
Cygnus X \citep{Schneider2006}. These are the DR21 range ($-$10-4
km s$^{-1}$), the W75N range (4-12 km s$^{-1}$), and emission
between 12 and 20 km s$^{-1}$ that we call the high-velocity (HV)
component.  The CO and \CII\ emission is mostly concentrated in bright
photodissociation regions (PDRs) among which the DR21 ridge and the
W75N cloud are well-known star-forming sites
\citep{Schneider2010,Hennemann2012}.  The bright \CII\ emission
features will be discussed elsewhere, here we focus on the low surface
brightness \CII\ emission in which the molecular clouds are embedded,
in particular on areas that appear devoid of CO emission in the W75N
and HV velocity ranges. Figure~\ref{fig1}e, f clearly shows that in
those regions where CO emission is lower than its 3$\sigma$ noise
level of 3.6 K km s$^{-1}$, \CII\ intensities are typically $\gtrsim$5
K km s$^{-1}$ (3$\sigma$ = 1.8 K km s$^{-1}$).  (See Methods and
Extended Data Fig.~\ref{intens-noise} for details).  We thus consider
these \CII-bright areas as CO-dark but recognize that there may be
faint CO emission below the detection limit. Substantial 
\CII\ emission at locations of CO-dark gas is also seen in individual
velocity channels, portrayed by a movie that scans through all
velocities. A one channel snapshot at +16.4 km s$^{-1}$ is shown in
Fig.~\ref{fig2movie}. The rather homogeneous \CII\ distribution argues
against an origin from photoevaporation flows from the surfaces of the
molecular clouds that would be more structured and intense toward the
clouds.

%

%
\begin{figure*}[ht]
\begin{center} 
\includegraphics [width=12cm, angle={0}]{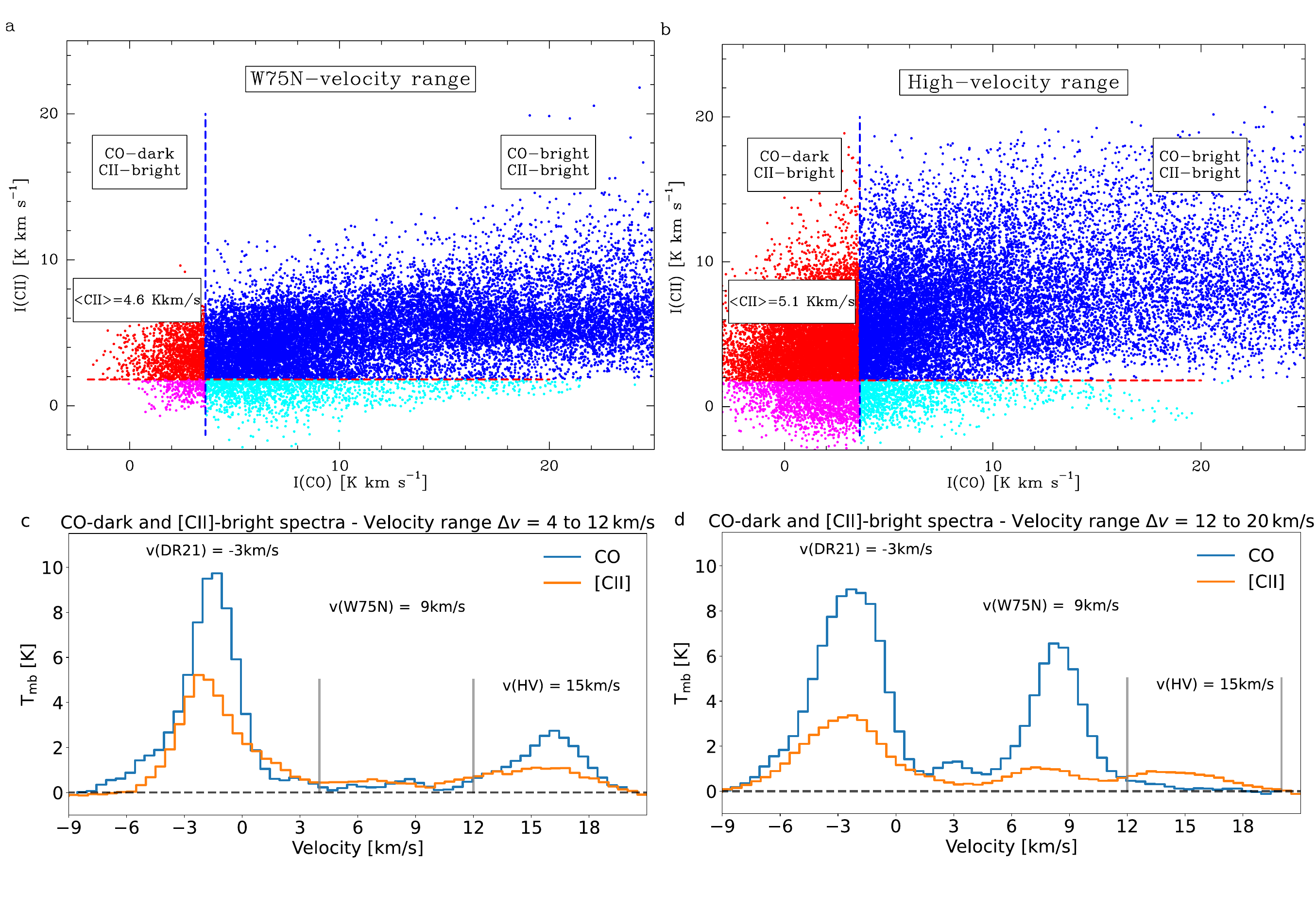}
\caption{Scatter-plots and spectra of \CII\ and CO 1$\to$0
  emission. {\bf a}, {\bf b}, The pixel-by-pixel correlations in the
  W75N (a) and HV (b) velocity ranges. The 3$\sigma$ noise levels for
  \CII\ emission (1.8 K kms$^{-1}$) and CO emission (3.6 K km
  s$^{-1}$) are indicated by a red and blue dashed line,
  respectively. The upper left area with red pixels indicates the
  intensity space that is below the CO noise level but bright in \CII,
  with an average value of 4.6 K km s$^{-1}$ (standard deviation 1.5 K
  km s$^{-1}$) and 5.1 K km s$^{-1}$ (standard deviation 2.2 K km
  s$^{-1}$) for the W75N and HV velocity ranges, respectively. {\bf
    c}, {\bf d}, The average spectra over all pixels in the map
  identified as CO-dark and \CII-bright in the scatter plot in the
  W75N (c) and HV (d) velocity ranges, indicated by grey vertical
  lines.  The r.m.s. noise of the spectra for both velocity ranges is
  0.075 K for \CII\ and 0.15 K for CO, respectively.  The \CII\ and CO
  line integrated emissions are 4.7 K km s$^{-1}$ and 2.2 K km
  s$^{-1}$ for the W75N range and 5.4 K km s$^{-1}$ and 1.3 K km
  s$^{-1}$ for the HV range, respectively.} \label{fig3scatter}
  \end{center} 
\end{figure*}

\begin{figure*}[ht]
  \begin{center} 
\includegraphics [width=13cm, angle={0}]{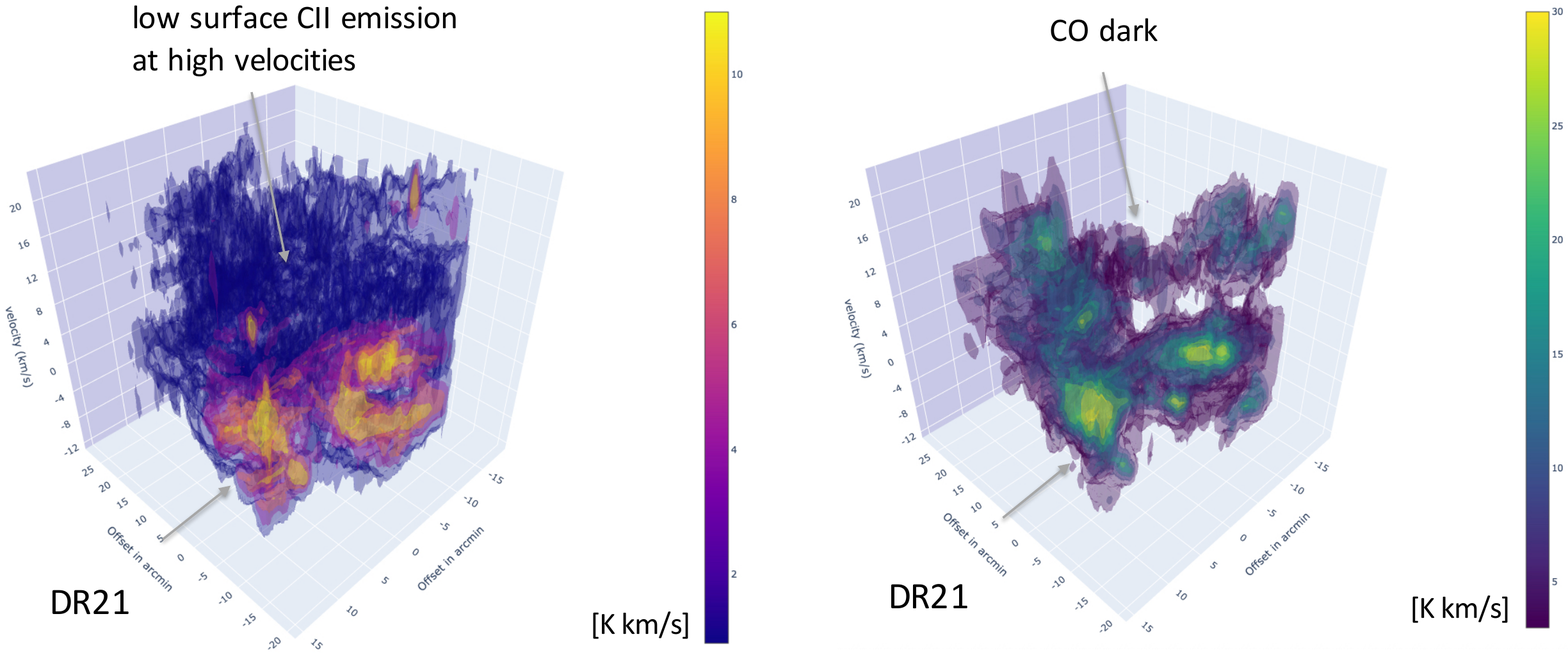}
\caption{3D iso-surface rendering of position-velocity cuts for
  \CII\ and CO 1$\to$0 emission over the entire area observed in
  \CII. The x- and y-axis are offsets in arcmin from the central map
  position, the z-axis is velocity in km s$^{-1}$. The emission starts
  at the 5$\sigma$ level for both tracers. The bright star forming
  cloud DR21 and other dense molecular clouds are embedded in a large
  scale cloud structure only visible in \CII\ (dark blue).  An
  interactive version of these plots is found at
  \url{https://astro.uni-koeln.de/stutzki/research/feedback/animations}. }
\label{pv-3d}
\end{center} 
\end{figure*}

From scatter plots (Figs.~\ref{fig3scatter}a and b) of \CII\ and CO
emission, we calculate a mean \CII\ intensity of $\sim$5 K km s$^{-1}$
in the CO-dark and \CII-bright regime and derive average \CII\ and CO
spectra (Fig.~\ref{fig3scatter}c and d) from these pixels (which are
obviously not the same for each velocity range). In total, 29\% (6\%)
of the area in the HV (W75N) range is CO-dark and \CII-bright,
compared to 63\% (88\%) for \CII\- and CO-bright gas. These values,
however, strongly depend on the total area that was mapped and are
subject to a selection effect because the Cygnus \CII\ mapping focused
on the bright PDR regions and less on the cloud outskirts. The
\CII\ spectra show a velocity bridge of emission between the clouds,
that is, DR21 at $-$3 km s$^{-1}$, W75N at 9 km s$^{-1}$ and the HV cloud
at +15 km s$^{-1}$. CO is also present, but clearly weaker, in
particular in the HV range. The kinematic connection in \CII\ becomes
particularly evident in three-dimensional (3D) position-velocity (PV) plots displayed for
\CII\ and CO in Fig.~\ref{pv-3d}.  The strongest emission in both
tracers is concentrated in the $-$3 km s$^{-1}$ component from the
DR21 ridge including the DR21 outflow and in the 9 km s$^{-1}$
component from the W75N cloud. These bright clouds and PDR regions
(yellow in the \CII\ image) are embedded in a pervasive medium
emitting in \CII\ (in dark blue) which is not or only little visible
in CO.

%

Summarizing, we conclude that instead of a head-on collision between a
$-$3 and +9 km s$^{-1}$ molecular cloud
\citep{Dickel1978,Dobashi2019}, we witness here an interaction of
several partly atomic flows (seen in \CII) and partly molecular flows
(seen in CO). The next section quantifies this scenario by calculating
the physical properties of the interacting gas.

%
\begin{figure}[ht]
\begin{center} 
\includegraphics[width=0.75\textwidth]{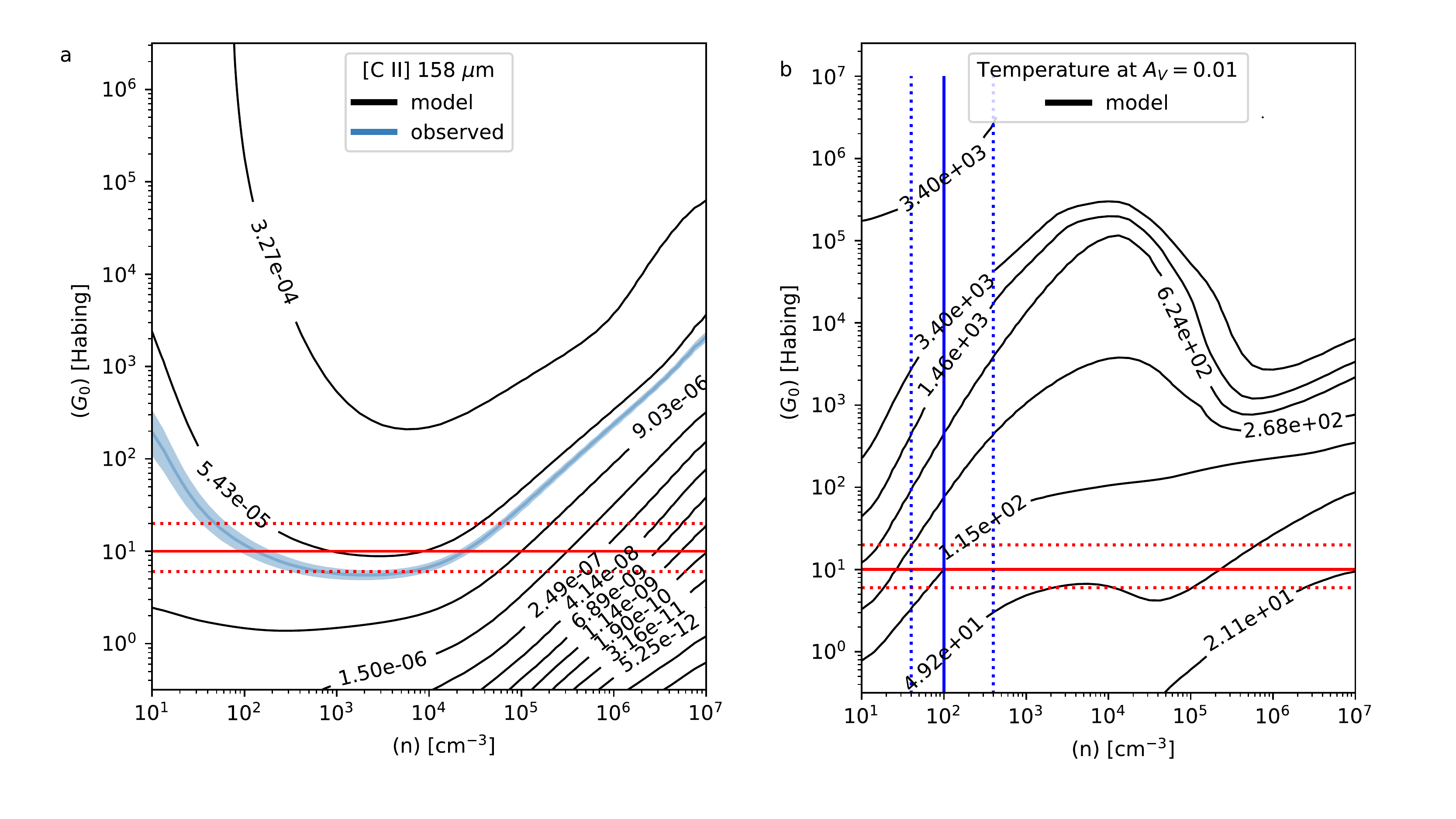}
\caption{PDR model results. The panels show the parameter space of
  hydrogen density $n$ and FUV field calculated from the PDR
  toolbox. {\bf a}, The dark blue isocontour of the observed
  \CII\ integrated intensity of 5 K km s$^{-1}$ and the noise r.m.s. of
  0.6 K km s$^{-1}$ in dashed light blue. The estimated FUV field of
  10 G$_\circ$ is indicated by a red line. The dotted red lines
  indicate approximately a FUV field of double and half the
  value. {\bf b}, The isocontours of surface temperature from the
  PDR model. For an average density of $\sim$100 cm$^{-3}$ (blue line)
  and a FUV-field of $\sim$10 G$_\circ$ (red line), we obtain a
  temperature of 115 K.  The dashed lines for density and temperature
  indicate how these values change if the FUV field is higher or
  lower.}
\label{pdr}
\end{center} 
\end{figure}

\subsection{Physical properties of the interacting gas}\label{gas}

We estimate the density and temperature of the gas detected with
\CII\ at velocities v $>$ 4 km s$^{-1}$ using predictions
\citep{Pound2023} from the PDR toolbox (Methods) for a \CII\ line
integrated intensity of 5 K km s$^{-1}$. From a census of the 169
OB-stars of Cyg OB2, we derive a Habing field of $\sim$10 G$_{\rm o}$
(Extended Data Fig. \ref{feld}), where G$_{\rm o}$ is the mean
interstellar radiation field.  The PDR model (Fig.~\ref{pdr}a)
indicates hydrogen densities of $\sim$100 cm$^{-3}$, which is typical
for diffuse gas at the transition from atomic to molecular.  We
exclude here the high-density solution ($>$10$^4$ cm$^{-3}$) because
then, significant CO emission should have been detected - which is not
the case. We note that all numbers have an uncertainty mostly because
of the adopted value of the FUV field. With the derived densities, we
obtain a surface temperature (Fig.~\ref{pdr}b) of 115 K for the PDR
gas layer.  This is an upper limit for the kinetic temperature
T$_{kin}$ of the gas, since the temperature drops entering deeper PDR
layers.  To narrow down T$_{kin}$, we performed a study of
\HI\ self-absorption (HISA) toward DR21 (Methods and Extended Data
Figs.\ref{hisa1} and \ref{hisa2}) and obtained a gas temperature of
$\sim$100~K.  We use this value to calculate C$^+$ and hydrogen column
densities, N(CII) and N(H), respectively (Methods and Extended Data
Fig.\ref{hisa3}), and give all input values and results in
Table~\ref{pdr-table}.  N(H) consists of an atomic and molecular part,
and the relative fractions are somewhat variable because the formation
of H$_2$ depends on the local radiation field and density, and on
turbulent mixing motions \citep{Bialy2017} that cause large- and
small-scale density fluctuations. We estimate (Methods) that
$\sim$23\% of the gas in the W75N range and $\sim$14\% in the HV range
is molecular. This is qualitatively in good agreement with results
from colliding \HI\ flow simulations \citep{Clark2019}, predicting
that about 20\% of hydrogen is in the form of H$_2$ at densities
around 100 cm$^{-3}$ for the initial phases of cloud formation. Our
values are also conform with the results of \citep{Franeck2018} who
find that $\lesssim$20\% of \CII\ comes from the molecular
phase. Their simulation set-up represents a section of the Milky Way
disc in which turbulence is injected by supernova explosions but the
dynamic effect of gas accretion on to the clouds from the larger
scale, galactic environment is retained. However, they investigate
only the earliest phases of cloud formation with an UV field of 1.7
G$_\circ$ and lower temperatures of $\sim$50 K.  The masses (Methods)
contained in the atomic gas are 7800 M$_{\rm sun}$ for the W75N range
and 9900 M$_{\rm sun}$ for the HV range, respectively. This is an important mass reservoir for building up more molecular clouds,
comparable to the fully molecular cloud DR21 ($\sim$15 000 M$_{\rm
  sun}$, \citep{Hennemann2012}). The time scale for cloud assembly is
given by the relative velocity of the components and their size. The
column densities of the W75N and the HV cloud translate into a size of
12~pc for a density of 100 cm$^{-3}$, leading to an assembly time of
1.3~Myr based on their separation in velocity space by about 10 km
s$^{-1}$. In a quasi-static scenario, molecular cloud formation would
take much longer, about 10~Myr at a density of 100 cm$^{-3}$, based on
the formation rate of molecular H$_2$ of $3 \times 10^{-17}$~cm$^3$
s$^{-1}$ \citep{Jura1974}. Faster cloud formation with significant
fractions of H$_2$ can be explained, however, from colliding flow
simulations that temporarily create pockets of gas with higher density
\citep{Clark2012}.

\begin{table}[h]
\begin{center}
\caption{CO-dark, \CII-bright gas physical properties}\label{pdr-table}
\begin{tabular}{cccccc}
  \hline
 I$_{\rm CII}$\footnotemark[1] & FUV field   & Density & Temperature & N(CII)\footnotemark[2] & N(H)\footnotemark[3] \\
 {\small [K kms$^{-1}$]}      & {\small [G$_{\rm o}$]} & {\small [cm$^{-3}$]} & [K] & {\small 10$^{18}$ [cm$^{-2}$]} & {\small 10$^{21}$ [cm$^{-2}$]} \\
   \hline
 $\sim$5                      & 10          & 100     & 100         &  0.606                &  3.78                   \\
\end{tabular}
\end{center}
\footnotesize{$^1$Average of the C$^+$ line integrated intensity in CO-dark regions. We call [CII]-bright all regions with [CII] emission above the 3$\sigma$ noise level.} \\ 
\footnotesize{$^2$The C$^+$ column density N(CII) is calculated with eq.~\ref{eq:NII} using the C$^+$ line integrated intensities.} \\ 
\footnotesize{$^3$The total hydrogen column density N(H) = N(HI) + 2 N(H$_2$) is estimated from N(CII), applying the abundance C/H = 1.6$\times$10$^{-4}$ \citep{Sofia2004}.}  
\end{table}


\section{Discussion} 

The observed levels of \CII\ intensity in the W75N and HV velocity
ranges (excluding bright local sources such as the W75N protocluster)
are consistent with the low FUV illumination of $\sim$10 G$_{\rm o}$
we estimated from the census of the stars from Cyg OB2 at a distance
of $\sim$1.6 kpc \citep{Apellaniz2022}. The HV gas can thus not stem
from the Cygnus Rift, a dark extinction feature at 600 pc with no
notable excitation sources \citep{Schneider2006}. Recently,
\citep{Lallement2022} used the GAIA2 data release in combination with
extinction to build 3D maps of the dust in the local arm and
surrounding regions and confirmed that there is no active
star-formation taking place in the Rift.

Maser parallax measurements \citep{Rygl2012} indicate that W75N
(1.3$\pm$0.1 kpc) is slightly but clearly in front of DR21
(1.5$\pm$0.1 kpc) showing that the densest parts of these two
molecular clouds could not have collided head-on.  Observed absorption
features in $^{12}$CO, HCO$^+$, CH$^+$, SH$^+$ toward DR21
\citep{Dickel1978,Schneider2010,Godard2012} support this 3D view (see
Methods for details on the Cygnus X complex). In addition, our HISA
study toward DR21 identifies broad absorption in the velocity range
$-$5 to 20 km s$^{-1}$ (Methods). Accordingly, the atomic clouds at
red-shifted velocities with v $>$ 4 km s$^{-1}$ (W75N, HV) must be
located in front of DR21 and the dynamics we traced in \CII\ indicates
that all three of them are clearly on collisional trajectories. Our
scenario of molecular cloud + \HI\ envelopes interaction, visible
through \CII, actually implies that the DR21, W75N and HV components
are not too far separated but should be located within a similar
volume with a radius of presumably 20-50 pc. More precise distance
estimates would help to test our view.

The composition of the gas seen in low surface brightness C$^+$
emission is about 20\% molecular and 80\% atomic. These values can be
compared with the findings from the GOTC+ survey
\citep{Pineda2013}. They observed Galactic sightlines with many bright
PDRs along the line-of-sight and derived that $\sim$47\% of \CII\ emission
arises from PDRs, $\sim$28\% from CO-dark gas, $\sim$21\% from cold
atomic gas, and $\sim$4\% from ionized gas. Our observations reveal a
large reservoir of CO-dark gas of several thousand solar masses, 
comparable in mass to the active star forming regions DR21 and W75N, 
that was previously unrecognized. We show that the DR21, W75N and HV
molecular clouds cross each other, interacting mostly through their
low-density enveloping atomic gas layers. The collision of small-scale
\HI\ flows forms dense, molecular gas in the compressed layer of
oblique shocks such as proposed in \citep{Inoue2018,Bonne2020b} but
the residual \HI\ is still there and forms a reservoir from which more
mass is accreted.  We expect that many of these compressed layers show
a flattened sheet-like structure, as increasingly seen in observations
\citep{Kabanovic2022,Pineda2022}, and work on a study to test this
scenario. We note that it depends strongly on the density if the
interacting gas is also seen in CO (at the same velocity as the
\CII). For the W75N range, we indeed observe a bridge of emission for
CO and \CII\ at the same velocities, most likely because the molecular
fraction is higher. Other studies \citep{Fukui2021}, only using CO,
already showed this molecular interaction, also at high velocities
\citep{Fukui2014}.  To a certain extent, it is also possible that the
reason we detect higher \CII\ velocities is that some of the CO is
already shocked due to the compression and thus at lower
velocities. The pre-shock high-velocity gas, however, is at low
densities and only visible in \CII.

Our scenario leads to a continuous assembly of more molecular material
on very short time scales over $\sim$ one million years.  It is not
likely, but cannot fully be excluded,  that the molecular clouds
formed in the collision zones of interacting expanding \HI\ shells
\citep{Inutsuka2015} because there are no clear observational
signatures for such \HI\ shells and the relative velocities we observe
in Cygnus are too large to be only driven by an expanding bubble.

We note that the magnetic field, and in particular its orientation,
can also play an important role in this sort of cloud
assembly. \citep{Inoue2009,Inoue2012} derived that a general alignment
of the collision axis and mean magnetic field direction is necessary
for direct cloud formation. Recently, \citep{Abe2022} showed that
under an inclination angle of 45$^o$, it is possible to accumulate
enough mass allowing the formation of massive clusters with O-type
stars. Alternatively the observed interaction between the \HI/CO
complexes have not formed directly the seeds of the CO clouds. Instead
we propose that the crossing of the \HI/CO clouds we observe have
triggered an acceleration of the concentration of these seeds into the
dense and massive star-forming clouds through the scenario proposed
for Musca by \citep{Bonne2020b}. For the quiescent Musca cloud,
\citep{Bonne2020b} found that molecular gas formed preferentially at
the convergence point of the magnetic field, bent behind the shock
front. For Cygnus X magnetic field measurements would thus be crucial
to address this important point.

The implications of such a generalized scenario to explain cloud
interactions, and consequently star formation, both in Musca and in
Cygnus, are important because they suggest a certain degree of
universality of cloud formation as a result of interactions between at
least partly diffuse \HI\ clouds. This is in line with the scenario of
the interplay of feedback bubbles and gravity leading to the formation
of dense structures in the multi-phase ISM \citep{Pineda2022}. The
main differences between Musca and Cygnus X are the larger initial
densities in Cygnus X and the velocity of the collision with $\sim$20
km s$^{-1}$ for Cygnus X and less than 10 km s$^{-1}$ in
Musca. Streams of diffuse gas with relative velocities of less than 10
km s$^{-1}$ are easily justified by turbulent \HI\ gas in the galaxy
since the sound speed in the warm neutral medium is of this order of
magnitude. The origin of gas streams of more than 20 km s$^{-1}$ is
more difficult to explain.  They can be driven by the complex
interplay between gravity and stellar feedback effects, and the
thermodynamic response of the multi-phase interstellar medium.  In
galaxy-wide simulations \citep{Dobbs2022} spiral density waves lead to
high-velocity collisions of flows that can form massive OB
clusters such as the ones seen in Cygnus X. In any case, our finding
will have major ramifications for our understanding of molecular cloud
assemblage in the Milky Way and other galaxies.  Further observations
of the extended \CII\ emission in the FEEDBACK sample will reveal if
this sort of interaction is common to other giant molecular cloud
regions. In the future, the GUSTO and ASTRHOS balloon projects will
measure the \CII\ emission in the Milky Way and in the Large and Small
Magellanic Clouds. A promising other tracer for detecting flows of
partly atomic, partly molecular gas is atomic carbon (\CI). It is
predicted to also trace the CO-dark gas component \citep{Clark2019}
and the \CI\ 1-0 line is observable from the ground. The GEco project
on the upcoming CCAT-prime/FYST telescope will perform extended
surveys in this line \citep{Simon2020}.




\section{Methods}\label{online}

\begin{figure*}[ht]
\begin{center} 
\includegraphics[width=0.6\textwidth]{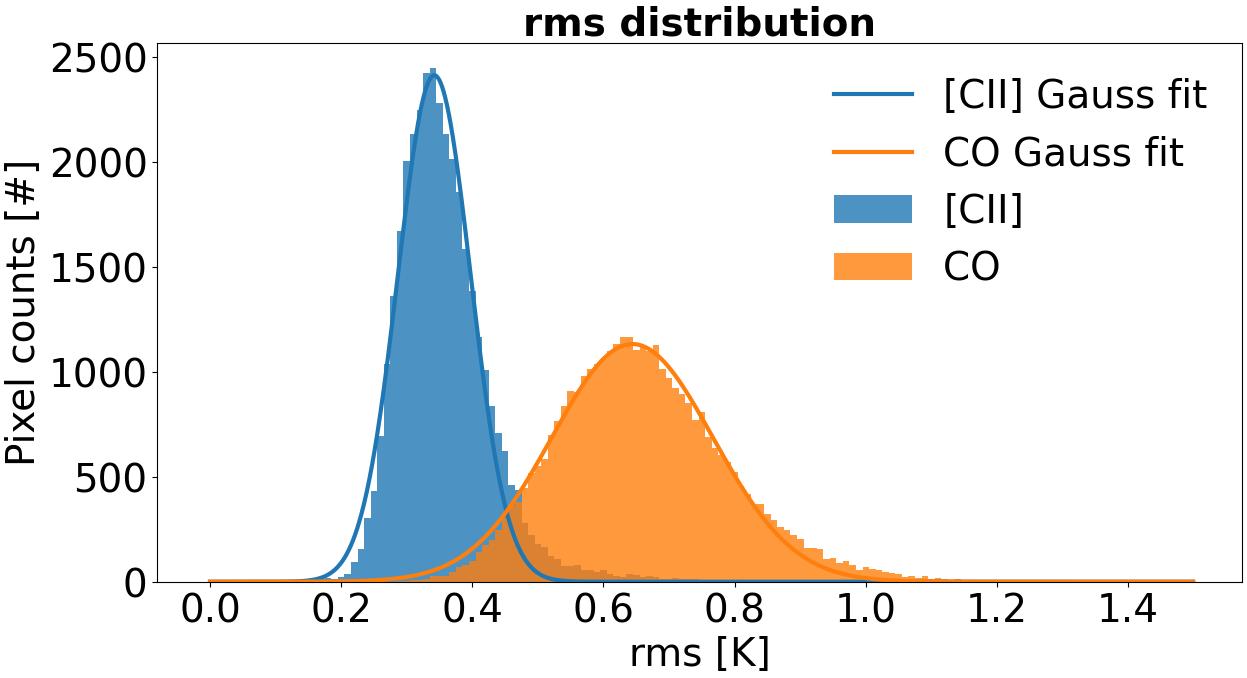}
\caption{Extended Data: R.m.s. noise distribution in 0.01 K bins for all
  map pixels in \CII\ (blue) and CO (orange). The peaks of the
  Gaussian distributions are at 0.34 K and 0.64 K, respectively.}
\label{noise}
\end{center} 
\end{figure*}

\begin{figure*}[ht]
\begin{center} 
\includegraphics[width=0.6\textwidth]{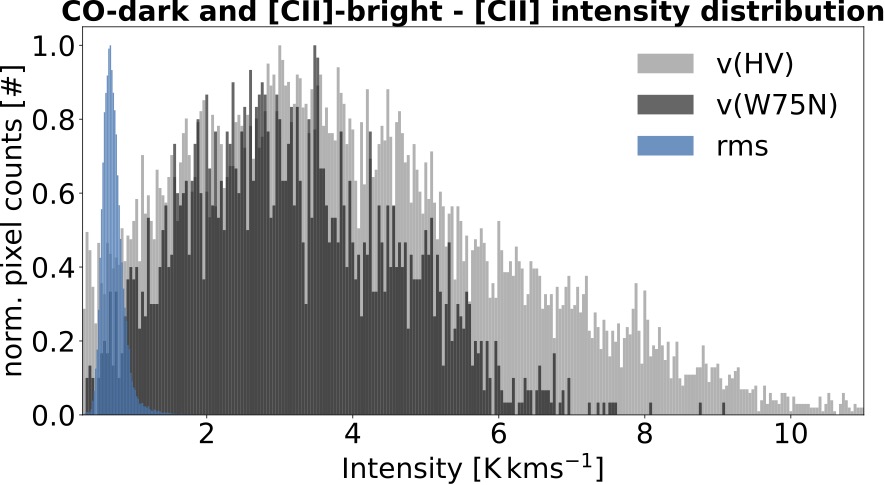}
\caption{Extended Data: Binned (0.02 K km s$^{-1}$ for the r.m.s. noise
  and 0.04 K km s$^{-1}$ for \CII, respectively) intensity normalized
  PDF (probability distribution function) for the \CII-bright, CO-dark
  map pixels together with the r.m.s. noise PDF. The W75N pixels are
  indicated in dark grey, the HV range in light grey and the r.m.s. in
  blue. }
\label{intens-noise}
\end{center} 
\end{figure*}

\subsection{Observations and data reduction} \label{methods:obs}
The Cygnus X region was observed during several flights in 11/2019,
2/2021, 11/2021 and 4/2022 for the legacy program FEEDBACK, using the
2.7m telescope onboard the Stratospheric Observatory for Infrared
Astronomy (SOFIA). The dual-frequency heterodyne array receiver
upGREAT \citep{Risacher2018} was tuned to the \CII\ 157.7~$\mu$m line
in the Low Frequency Array (LFA) 2$\times$7 pixel array and the
\OI\ 63~$\mu$m line (not shown) was observed in the High Frequency 
7 pixel array (HFA).  The LFA has an array size of 72.6$''$, and
a pixel spacing of 31.8$''$. The half-power beam width at 158~$\mu$m
is 14.1$''$, determined by the instrument and telescope optics, and
confirmed by observations of planets. For each flight series, the
main-beam efficiencies $\eta_{mb}$ for each pixel for the LFA and HFA
channels were determined, the average value for the LFA is 0.65 and we
use this value here.  The entire map region was split into multiple
square "tiles" with 435.6$''$ on one side and each square was covered
4 times. The on-the-fly (OTF) scan speed was selected to attain Nyquist sampling of
the LFA beam (dump every 5.2$''$). The total time for one OTF line was
25.2 s, that is together with the OFF observation, within the measured
Allan variance stability time of the system.  The first two coverages
were done once horizontally and vertically with the array rotated
19$^\circ$ against the scan direction, so that scans by 7 pixels are
equally spaced. The second two coverages were then shifted by 36$''$
in both directions to achieve the best possible coverage for the
\OI\ line in the LFA array mapping mode.  Overall, each tile took
$\approx$50 min. to complete. The horizontal and vertical scanning can
cause some striping in the maps. In order to reduce these effects, we
applied a Principal Component Analysis (PCA) on the data.  In short,
our method based on PCA uses the information of systematic variations
in the baseline from a large set of observations from the
emission-free OFF-position that are regularly taken during each
scan. These variations are caused by time-dependent instabilities in
the backends, receiver, telescope optics, and atmosphere.  We produce
an “OFF–OFF” spectra by subtracting subsequent OFF positions from each
other and calibrate the data in the same way as the “ON–OFF” spectra
that contain the astronomical source emission. We then identify
systematic “eigenspectra” in the OFF$-$OFF spectra that account for
all or at least most of the structure in the baseline. Using a linear
combination of the strongest components, we reconstruct the ON–OFF
spectra with the best-fit coefficients for each component. We scale
each component by the coefficients that were found and subtract those
from the ON–OFF spectra. This procedure removes the systematic
variations found in the OFF–OFF spectra but does not make changes to
the astronomical line in terms of intensity, width, position etc. in
the ON–OFF spectra.  We further improve the data quality by using a
sophisticated algorithm to determine a set of spectra that are free of
emission from the PCA-corrected spectra for each frontend and SOFIA
flight. With this set of emission free spectra we employ a second
PCA-correction equal to the one described above, but using now these
emission-free spectra to determine the baseline "components". With
these components we correct systematic variations that happen on
timescales of the individual OTF-dumps, which are much shorter than
the 10\,s integration time for OFF spectra.  The final PCA map was
then compared to one obtained by removing a polynomial baseline of
order 3, performing difference maps, ratio maps, scatter plots
etc. and we found no systematic effects.

We then spatially smoothed the resulting \CII\ map to an angular
resolution of 30$''$ in a Nyquist sampled 10$''$ grid in order to
emphasize the large-scale emission distribution and not focus on
small-scale variations.  Some striping effects are still visible, but
they do not produce systematic effects in the data analysis (scatter
plots for example).  The map center position is at RA(2000) =
20$^h$38$^m$39.3$^s$, Dec(2000) = 42$^\circ$20$'$39.3$''$ and an
emission-free OFF position at RA(2000) = 20$^h$39$^m$48.34$^s$,
Dec(2000) = 42$^\circ$57$'$39.11$''$ was used.  A Fast Fourier
Transform Spectrometer (FFTS) with 4 GHz instantaneous bandwidth
\citep{Klein2012} with a velocity resolution of 0.04 km s$^{-1}$ (from
the hardware selected frequency resolution of 0.244 MHz) served as
backend. We here employ data resampled to 0.5 km s$^{-1}$ resolution.
For further technical details see \citep{Schneider2020}.  All spectra
are presented on a main beam brightness temperature scale T$_{mb}$,
i.e., corrected for the main beam efficiency.  The r.m.s. noise of this
final data set (and the one of CO emission) was then determined by
fitting a 0th order baseline to each spectrum, excluding the windows
with emission. The noise is homogeneous across the maps and follows
approximately a Gaussian distribution (Extended Data Fig.~\ref{noise})
with a peak at 0.34 K for \CII\ and 0.64 K for CO, respectively. As a
second method, we determined the noise by spatially averaging the
pixels in emission free channels and obtained values of 0.31 K for
\CII\ and 0.62 for CO, respectively.  For the line integrated
intensities, we calculated the $\sigma$ noise level for the 8 km
s$^{-1}$ wide velocity integration range by $\sigma = \sqrt(16) \times
0.6 \times 0.5$ = 1.2 K km s$^{-1}$ in which 0.6 K is the noise, 0.5
km s$^{-1}$ is the channel width and 8 km s$^{-1}$ correspond to 16
channels. The 3$\sigma$ CO level is thus 3.6 K km s$^{-1}$.  The
equivalent calculation for \CII\ delivers a 3$\sigma$ level of 1.8 K
km s$^{-1}$. In addition, we show in Extended Data
Fig.~\ref{intens-noise} a PDF of the \CII\ bright, CO-dark intensities
together with a PDF of the r.m.s. noise to demonstrate that the observed
W75N and HV intensities are clearly offset from the noise. We also
note that the smoothing is the same for all data and cannot change any
structural behaviour.

A potential problem could be that a part of the emission that we find
in the W75N velocity range might stem from a \TCII\ hyperfine
transition at DR21 source velocities.  The \TCII\ transition splits
into three hyperfine components with a relative strength of
$s_{\mathrm{2\to 1}} = 0.625$, $s_{\mathrm{1\to 0}} = 0.25$ and
$s_{\mathrm{1\to 1}} = 0.125$ caused by the unpaired spin from the
additional neutron. The three satellites are velocity shifted $\Delta
{\rm v}_{\mathrm{2\to 1}} = 11.2 \,\mathrm{km \, s^{-1}}$, $\Delta
{\rm v}_{\mathrm{1\to 0}} = -65.2\,\mathrm{km \,s^{-1}}$ and $\Delta
{\rm v}_{\mathrm{1\to 1}} = 63.2\,\mathrm{km \,s^{-1}}$ with respect
to the \CII\ fine structure line \citep{Ossenkopf2013}.  The
F$_{2\to1}$ component of \TCII\ emission from the bright PDRs at DR21
velocities ($-$3 km s$^{-1}$) may thus appear at $\sim$8 km s$^{-1}$,
close to the systemic velocity of W75N ($\sim$9 km s$^{-1}$).  We
estimated this contribution by comparing the \CII\ emission in the
DR21 velocity range with the emission in the W75N velocity range for
the \CII-bright, CO-dark gas and found no correlation between the two
quantities at a DR21 \CII\ line strength of $\sim$15 K km s$^{-1}$
(note that the average \CII\ intensity is 5 K km s$^{-1}$). For
optically thin \CII\ this translates into an expected
\TCII\ F$_{2\to1}$ line strength of 0.16 K km s$^{-1}$. Even if
optical depth effects may increase this value somewhat, it is so small
compared to our noise limit that it certainly does not change our
quantitative estimates and simultaneously explains that there is no
correlation detected.

\subsection{Photodissociation region modelling} \label{methods:pdr}

We employ the observed \CII\ intensities using the plane-parallel
models provided by the PDR toolbox found at
\href{https://dustem.astro.umd.edu}{https://dustem.astro.umd.edu}
\citep{Pound2023}.  In short, these models solve the radiative
transfer equation with chemical balance and thermal equilibrium for a
plane-parallel PDR layer exposed to a UV radiation field, cosmic-rays,
and soft X-rays incident on one side.  A given set of gas phase
elemental abundances and grain properties is calculated, and the
emergent \CII\ line integrated intensities as a function of density
$n$ and radiation field G$_{\rm o}$ in Habing units are given. The
beam filling is assumed to be unity, which we consider a good approach
for the W75N and HV emission because the \CII\ arises from extended,
mostly diffuse gas. We here use the model WK2020 that features some
updates of the photorates and dependence with PDR depth, $^{13}$C
chemistry and line emission, and O collision rates. However, there is
nearly no difference in the \CII\ model prediction compared to the
2006 model.  The specific parameters for the WK2020 model are listed
in Table 1 in \citep{Pound2023}.  Some important values are the cosmic
ray ionization rate per H nucleus of 2 10$^{-16}$ s$^{-1}$ and the
formation rate of H$_2$ on dust of 6 10$^{-17}$ s$^{-1}$.

We focus here on modelling only the \CII\ emission though the
plane-parallel PDR model also predicts a CO (1$\to$0) brightness.
However, this value is very sensitive to the assumed total depth of
the cloud. The \CII\ intensity and the surface temperature trace
surface properties while the CO emission only arises from the layers
deeper in the cloud where CO can form, this means those that are not
CO-dark. As the toolbox models assume a gas column of visual
extinction Av = 7, implying a large fraction of CO-bright gas, while
the observed gas rather has a column of 3.8 10$^{21}$ cm$^{-2}$ we
expect, and see, a significant over-prediction of the CO 1-0
intensity by the model.

Considering the \CII\ intensity, we derive a density of n=100
cm$^{-3}$ for a FUV field of 10 G$_\circ$. Taking into account the
uncertainties in the FUV field, we assume a field that is double
($\sim$20 G$_\circ$) and half ($\sim$5 G$_\circ$) of the average value
and derive then a density range of $\sim$40 to $\sim$400
cm$^{-3}$. The corresponding surface temperatures are then $\sim$200 K
for n=40 cm$^{-3}$ and $\sim$90 K for n=400 cm$^{-3}$,
respectively. From our HISA study, however, we obtain a temperature of
90-120 K with an average of 108 K, which points towards a density of
$\sim$100 cm$^{-3}$ and thus to a FUV field of 10 G$_\circ$. A
significantly lower UV field would push the densities towards higher
values of more than 10$^3$ cm$^{-3}$, but we note that the observed
\CII\ isocontour is a very flat curve for high densities. In any case,
we would have detected CO emission in densities above 10$^3$
cm$^{-3}$. An even higher UV field has a smaller impact on the
density, but would still move values below $\sim$40 cm$^{-3}$. For
simplicity, we will use a common temperature of 100 K and a density of
100 cm$^{-3}$, but work out the values for column densities fractions
using these extreme limits of density and temperature.

\subsection{Calculation of physical properties} \label{methods:prop}

We assume that the major collision partner for C$^+$ is atomic hydrogen, 
and that the \CII\ line is optically thin, so that the line emission must be
sub-thermally excited, in view of the low densities. 
The \CII\ column density N(CII) is then calculated
\citep{Goldsmith2012} from
\begin{equation}  \label{eq:NII} 
{\rm N(CII)}  =  \frac{{\rm I}_{\rm CII} 10^{16}}{3.43} \times \left[  1 + 0.5 \times \exp(\frac{91.25}{{\rm T}_{kin}}) (1+\frac{2.4 \times 10^{-6}}{C_{ul}}) \, [{\rm cm}^{-2}] \right]
\end{equation}
with the line integrated \CII\ emission I$_{\rm CII}$ in [K km s$^{-1}$], the kinetic temperature T$_{kin}$ [K], and the
de-excitation rate $C_{ul}$ [s$^{-1}$] 
\begin{equation}  \label{eq:Cul} 
C_{\rm ul} =  n \, \times \, R_{ul}    
\end{equation}
with hydrogen density $n$ [cm$^{-3}$] and de-excitation rate coefficient $R_{ul}$, derived with  
\begin{equation}  \label{eq:Rul} 
R_{\rm ul} =  7.6 \times 10^{-10} \mathrm{cm^3 s^{-1}} ({\rm T}_{kin}/100)^{0.14}.  
\end{equation}
The total hydrogen column density N(H) = N(HI) + 2 N(H$_2$) is
estimated from N(CII), assuming all carbon is in the form of C$^+$ and
applying the abundance C/H = 1.6$\times$10$^{-4}$
\citep{Sofia2004}. For the nominal values of T=100 K and n=100
cm$^{-3}$, we obtain N(CII)=0.61 10$^{18}$ cm$^{-2}$ and N(H)=3.78
10$^{21}$ cm$^{-3}$. This total hydrogen column density corresponds
very well to the one estimated from our HISA study, assuming that most
of the gas seen in \CII\ is atomic. We thus consider a density of
$\sim$100 cm$^{-3}$ as the most likely value for the atomic
gas. Nevertheless, with a lower density of n=40 cm$^{-3}$ and T=200 K,
we calculate N(CII)=0.85 10$^{18}$ cm$^{-2}$ and N(H)=5.31 10$^{21}$
cm$^{-2}$. With n=400 cm$^{-3}$ and T=90 K, we derive N(CII)=0.20
10$^{18}$ cm$^{-2}$ and N(H)=1.22 10$^{21}$ cm$^{-2}$.

The mass of the atomic gas is then estimated by 
\begin{equation}  \label{eq:mass} 
M =  \frac{{\rm N}_{\rm CII} \, A \, m_H}{C/H}   [M_{\rm sun}]   
\end{equation}
with the area $A$ in cm$^{2}$, the mass of hydrogen $m_H$ in kg and the C/H abundance 1.6$\times$10$^{-4}$ \citep{Sofia2004}.

\subsection{Determination of the molecular fraction} \label{methods:frac}

The diffuse gas at high velocities ($>$4 km s$^{-1}$) is partly molecular and partly atomic. We here give a rough estimate of the
molecular fraction, which is defined \citep{Goldsmith2012} by
\begin{equation}
f({\rm H}_2) \, = \, \frac{2 {\rm N}({\rm H}_2)}{{\rm N}(\rm {H})}.   
\end{equation}

The molecular hydrogen column density N(H$_2$) is derived using the
average CO line intensities (I$_{{\rm CO}}$) from the spectra in the
W75N velocity range (2.2 K km s$^{-1}$) and the HV range (1.3 K km
s$^{-1}$). Both values have an error of 1.2 K km s$^{-1}$.  With a
CO-to-H$_2$ conversion factor X$_{{\rm CO}}$ = N(H$_2$)/I$_{{\rm CO}}$
of 2$\times$10$^{20}$ cm$^{-2}$ (K km s$^{-1}$)$^{-1}$, commonly used
in the literature, we obtain H$_2$ column densities of 0.44 10$^{21}$
cm$^{-2}$ and 0.26 10$^{21}$ cm$^{-2}$ for the W75N and HV velocity
range, respectively.  The total hydrogen column density N(H) is
derived from the \CII\ column densities, given in
Table~\ref{pdr-table}, with a value of 3.78 10$^{21}$ cm$^{-2}$ in
each velocity range.  With these values, the molecular fraction is
then calculated to be $f({\rm H}_2)$ = 0.23 for the W75N velocity
range and $f({\rm H}_2)$ = 0.14 for the HV range, respectively, with
an average value of $f({\rm H}_2)$ = 0.19. The portion of molecular
gas is thus nearly twice as much in the W75N velocity range compared
to the HV range. We note that the H$_2$ column densities and molecular
fractions are lower limits since we use the canonical value of the
X$_{{\rm CO}}$ factor which is mostly valid for evolved molecular
clouds.  In case the atomic gas has a lower or higher density because
of a different incident FUV field, the molecular fractions change
accordingly. We estimate that for n=40 cm$^{-3}$, $f({\rm H}_2)$ =
0.17 for the W75N velocity range and $f({\rm H}_2)$ = 0.10 for the HV
range, respectively. In the high density case with n=400 cm$^{-3}$,
$f({\rm H}_2)$ = 0.72 for the W75N velocity range and $f({\rm H}_2)$ =
0.43 for the HV range, respectively.  These values are more extreme
and less likely than the fractions we obtained with the nominal values
but we note that all observational and model values have their
uncertainties.

\begin{figure}[h]
\centering
\includegraphics[width=0.85\textwidth]{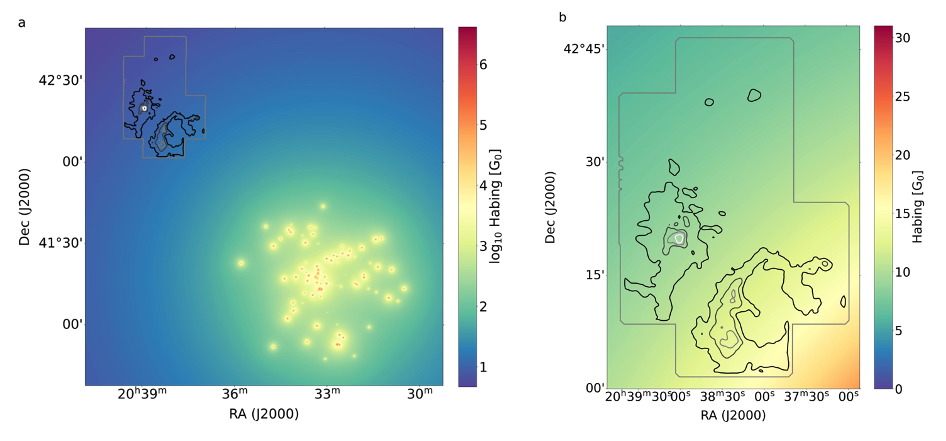}
\caption{Extended Data: Far-UV field in the Cygnus region. {\bf a} The
  large-scale FUV field on a log-scale in Habing units G$_{\rm o}$,
  determined from a census of the O and B-stars from the Cyg OB2
  cluster that are indicated in the panel. The grey box outlines the
  area shown in panel b. The observed \CII\ emission is overlaid with
  black contours corresponding to 50 K km s$^{-1}$ to 210 K km s$^{-1}$ by 40 K km s$^{-1}$.  
  {\bf b} The FUV field in the DR21 and W75N region on a linear scale with the \CII\ contours identical
  to the ones in panel a. }\label{feld}
\end{figure}

\subsection{Evaluation of the FUV field} \label{appA} 

The FUV field in Cygnus X was derived from a census of the stars of
Cyg OB2, using the compilation of \citet{Wright2015}. They listed 169
OB stars including 52 O-type and 3 Wolf-Rayet stars. To determine the UV-luminosity of the cluster, 
we assume that the spectral radiance of each star can be represented by a black-body:

\begin{equation}
    B\left(\lambda, T\right) = \frac{2hc^2}{\lambda^5}\frac{1}{e^{\frac{hc}{\lambda k_{\rm B} T}-1}}~,  
\end{equation}

with the wavelength $\lambda$, the speed of light $c$, the
Boltzmann-constant $k_{\rm B}$, the Planck-constant $h$ and the
temperature of each star $T$. To extract the UV portion of the
luminosity $L$, we integrate the Planck function over the UV range
between $910\,\AA$ and $2066\,\AA$, which corresponds to a photon
energy range of 6 to $13.6\,{\rm eV}$. The ratio of the integrated
spectral radiance over the UV range to the entire black-body spectrum
gives us the UV luminosity:

\begin{equation}
    L_{\rm UV} = \frac{\pi \int\limits_{\lambda_{910}}^{\lambda_{2066}} B\left(\lambda, T\right)\,\rm{d}\lambda}{ \sigma T^4} L~,
\end{equation}

with the Stefan-Boltzmann constant $\sigma$. The superposition of the
stellar UV flux of all stars considered gives the UV-field at every
point (RA, Declination) of the grid:

\begin{equation}
    F_{\rm UV} = \sum\limits_{i} \frac{L_{\rm{UV},i}}{4\pi R^2_{i}}~,
\end{equation}

where $R_i$ is the radial distance to each star.  We assumed the most
recent distance estimate, based on GAIA, of $1.6\,{\rm kpc}$
\citep{Apellaniz2022} for every star in the cluster to the observer
though there can be line-of-sight distance differences between the
individual stars. Extinction of the UV field by gas of the
interstellar medium is not taken into account, thus the determined UV
field is an upper limit.  The resulting FUV field is shown in Extended
data Fig.~\ref{feld} where it becomes obvious that the FUV field
varies between $\sim$5 and 20 G$_{\rm o}$ across the \CII\ map.  We
use a value of 10 G$_{\rm o}$ for our PDR modelling.  We note that a
closer distance of Cyg OB2 of $\sim$1.45 kpc \citep{Hanson2003} would
increase the UV field to values between $\sim$10 and 30 G$_{\rm o}$
with little influence on the PDR modelling.

\begin{figure*}[ht]
\begin{center} 
\includegraphics[width=0.85\textwidth]{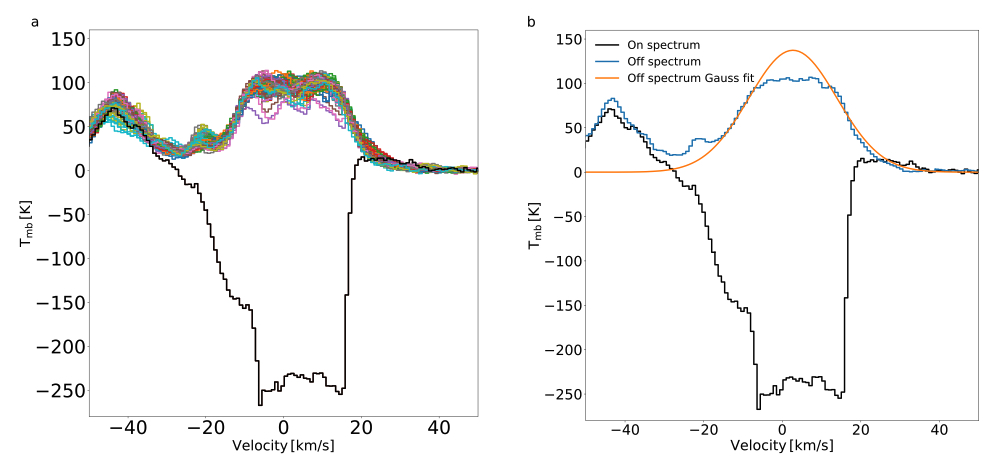}
\caption{Extended Data: \HI\ spectra toward DR21. {\bf a} The \HI\ absorption
  spectrum is shown in black, the 100 colored spectra represent the
  \HI\ emission within a square of $20'\times 20'$ around DR21 in a
  grid of 1$'$. {\bf b} The black curve is again the absorption
  \HI\ spectrum toward DR21, the blue curve is the \HI\ emission
  spectrum (off), and the orange curve shows the Gaussian fit to the
  blue off spectrum.}
\label{hisa1}
\end{center} 
\end{figure*}

\begin{figure*}[ht]
\begin{center} 
\includegraphics[width=0.65\textwidth]{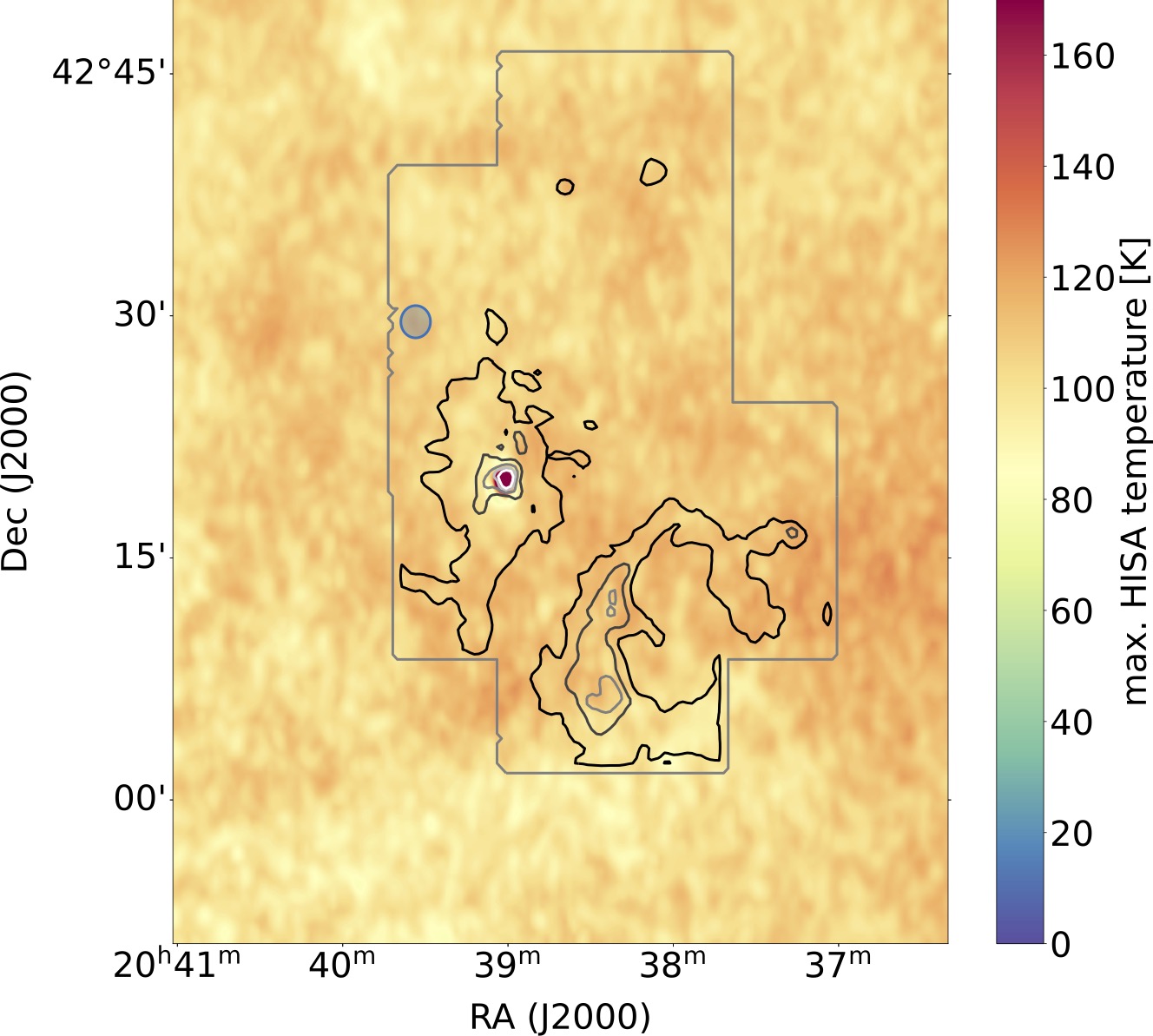}
\caption{Extended Data: Maximum possible HISA temperature
  around DR21. The strong continuum source DR21 stands out as a red spot with high temperatures
  of $\sim$175 K. Overall, the HISA temperature ranges between
  $\sim$90 and 120 K with an average of $\sim$100 K. The overlaid
  \CII\ contours go from black to white with six contour levels
  corresponding to 50 K km s$^{-1}$ to 210 K km s$^{-1}$ by 40 K km
  s$^{-1}$. The blue filled circle indicates the location of the off
  position. }
\label{hisa2}
\end{center} 
\end{figure*}

\begin{figure*}[ht]
\begin{center} 
\includegraphics[width=0.65\textwidth]{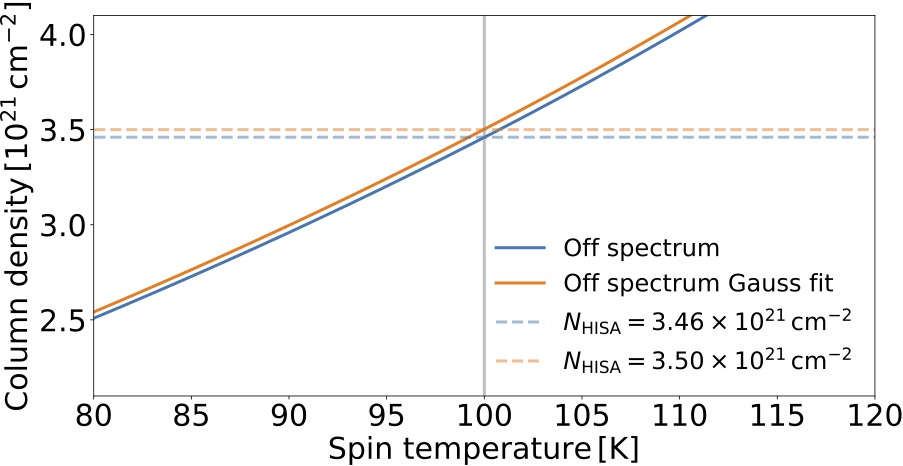}
\caption{Extended Data: HISA column density as a function of temperature. The solid
  blue curve shows the results for the off position and the solid
  orange curve for the Gaussian fit to the off position. The blue and
  orange dashed lines indicate the column density at $100\,{\rm K}$
  for the off position and the Gaussian fit, respectively.}
\label{hisa3}
\end{center} 
\end{figure*}

\subsection{\HI\ self-absorption} \label{appB}    
%
Studying self-absorption of atomic hydrogen toward a strong continuum
source is a way to determine the atomic column density in front of the
source. We follow the approach of \citep{Kabanovic2022} to derive the
amount of cold absorbing \HI\ in front of DR21, which emits strongly
in the 1.4 GHz continuum. Both, the \HI\ 21cm line data and the
continuum, stem from the Canadian Galactic Plane survey
\citep{Taylor2003} and have an angular resolution of 1$'$. First, the
method requires to estimate the \HI\ emission by finding a
mostly absorption free position (off) as a reference. For that, 100
\HI\ emission spectra around DR21 within a $20'\times 20'$ square were
plotted (Extended Data Fig.~\ref{hisa1}a) and compared to the average
\HI\ spectrum towards DR21, showing absorption across a large velocity
range. We are mostly interested in the range v = 4 to 20 km
s$^{-1}$. All off spectra show a similar line profile with a flat-top
shape peaking at a temperature of $\sim$100 K at the absorption dip of
the on spectrum. However, the various small absorption dips in the off
spectra indicate that the absorbing hydrogen cloud is also extended.  
As a best candidate we chose a spectrum that shows the least absorption features. The
location of this off spectrum is plotted in Extended Data Fig.~\ref{hisa2},
indicated by a blue circle of radius 1$'$, having a central position 
right acension(J2000) = 20$^h$39$^m$34.07$^s$, Declination(J2000) = 42$^\circ$29$'$38.00$''$.
The off spectrum is shown in Extended Data Fig.~\ref{hisa1}b in
blue, together with a Gaussian fit optimized to fit the wings of the
off spectrum. It could represent the \HI\ emission if some emission is
still absorbed at our off position. We calculated the velocity
resolved optical depth for both choices, the off spectrum and the
fitted one. This helps us to assess the uncertainties in the
derivation of the column density. The optical depth as a function of
velocity v is given by

\begin{equation}
    \tau_{\mathrm{HISA}}({\rm v}) = -\ln\left(1-\frac{T_{\mathrm{on-off}}({\rm v})}{T_{\mathrm{HISA}} -T_{\mathrm{off}}({\rm v})-T_{\mathrm{cont}} } \right)
    \label{eq:tau_HISA}
\end{equation}

\noindent
with the continuum temperature $T_{\mathrm{cont}}$ (437 K as an
average) for a HISA temperature $T_{\mathrm{HISA}}$. The corresponding
column number density of the absorbing \HI\ layer,
N$_{\mathrm{HISA}}$, can be determined by integration over
$\tau_{\mathrm{HISA}}({\rm v})$:

\begin{equation}
    \frac{N_{\mathrm{HISA}}}{\mathrm{cm^{-2}}} = 1.8224\cdot10^{18}\frac{T_{\mathrm{HISA}}}{\mathrm{K}}\int\tau_{\mathrm{HISA}}({\rm v})\,d\left( \frac{{\rm v}}{\mathrm{km/s}} \right)~.
\end{equation}

We integrate over the velocity range of 4 km s$^{-1}$ to 20 km
s$^{-1}$. The temperature dependent HISA column number density is
shown in Extended data Fig.~\ref{hisa3} for both, the off position and the Gaussian
fit. It becomes obvious that the differences are small.  To further
constrain the possible range of HISA column density, we determine the
maximum HISA temperature by the temperature at the absorption dip
minimum at each spectrum of the \HI\ map:

\begin{equation}
    T_{\mathrm{HISA,max}} = T_{\mathrm{on, min}} + T_{\mathrm{cont}}~.
\end{equation}

\noindent
The resulting temperature map is shown in Extended data
Fig.~\ref{hisa2} and confirms what we already concluded from Extended
data Fig.~\ref{hisa1}, that is, that the HISA temperature can not be
higher than $\sim 100\,{\rm K}$, which results in a HISA column
density of $\sim$N$_{\mathrm{HISA}}\sim 3.5\cdot10^{21}\,{\rm
  cm^{-2}}$. For a density of 100 cm$^{-3}$, this translates into a
\HI\ layer of 11 pc which compares well to other dominant molecular
structures in Cygnus X, like for example the DR21 ridge with a size of
$\sim$7 pc.

\subsection{The Cygnus X star-forming complex and the distance problem} \label{appC} 

The Cygnus X star forming region has for a long time been noticed as
an outstanding area because it lies around l = 90$^{\rm o}$ where the
local Galactic arm, the Perseus arm, and the outer Galaxy are found
along the same line of sight, covering distances between 1 and 8 kpc.
Since radial velocities around the tangent point in Cygnus X are close
to zero, they do not provide reliable distances. This is why the
Cygnus X region was long proposed to be an accumulation of clouds at
various distances along the different spiral arms
\citep{Piepenbrink1988}.  However, based on CO data and arguments of
UV-illumination, \citep{Schneider2006} proposed that most of the
observed velocity components seen in CO were associated and part of a
single complex despite the very large differences in velocities ($\sim
-$5 km s$^{-1}$ to 18 km s$^{-1}$). The main molecular cloud regions
are DR21 (around $-$3 km s$^{-1}$) and W75N (around 9 km
s$^{-1}$). At that time, it was not clear how such large relative
velocities could co-exist spatially and inside a single event of star
formation.  The scenario of a single complex for Cygnus X
\citep{Schneider2006} has been confirmed by maser parallax
measurements \citep{Rygl2012}, obtaining roughly the same distance for
DR21 and W75N despite their velocity difference of $\sim 12\,$km
s$^{-1}$. Our study now shows that the W75N and the HV
atomic/molecular clouds are located in front of DR21 (\CII\ emission
and \HI\ absorption) and interact with each other.


\subsection{Data availability} \label{methods:data}
The calibrated and polynominal baseline-corrected \CII\ data cubes at a velocity resolution of 0.2 km s$^{-1}$ are provided by the IRSA/IPAC archive and are found at
\href{https://irsa.ipac.caltech.edu/applications/sofia}{https://irsa.ipac.caltech.edu/applications/sofia}
under the project Number 07\_0077 (FEEDBACK, PIs are A.G.G.M. Tielens and N. Schneider). The PCA reduced data set will also be made available on the
IRSA/IPAC archive together with other data sets from the FEEDBACK program as level 4 data products. 

\subsection{Code availability} 
The results generated in this work are based on publicly available software
packages such as Python and do not involve the extensive use of custom code.

\subsection{Acknowledgments}
This study was based on observations made with the NASA/DLR
Stratospheric Observatory for Infrared Astronomy (SOFIA). SOFIA is
jointly operated by the Universities Space Research Association
Inc. (USRA), under NASA contract NNA17BF53C, and the Deutsches SOFIA
Institut (DSI), under DLR contract 50 OK 0901 to the University of
Stuttgart. upGREAT is a development by the MPIfR and University of
Cologne, in cooperation with the DLR Institut f\"ur Optische
Sensorsysteme. \\
Financial support for FEEDBACK at the University of Maryland was provided by NASA through award SOF070077 issued by USRA. \\
The FEEDBACK project is supported by the BMWI via DLR, Projekt Number
50 OR 1916 and 50 OR 2217. \\
N.S., S.B., R.S., and L.B. discloses support through the project "GENESIS" from the funder 
ANR-16-CE92-0035-01/DFG1591/2-1. \\
This work was supported by the German DFG/CRC project number SFB 956.\\
The research presented in this paper has used data from the Canadian Galactic Plane Survey, a Canadian project with international partners, supported by the Natural Sciences and Engineering Research Council. \\
L.B. was supported by a USRA postdoctoral fellowship, funded through the NASA SOFIA contract NNA17BF53C.

\subsection{Author contributions} 
N.S. and A.G.G.M.T. are the principal investigators (PIs) of the FEEDBACK project and prepared the SOFIA proposal. 
J.S. is the PI of the GREAT instrument, M.M. and O.R. are GREAT instrument scientists.   
R.S., C.B., and N.S. (plus other members of the GREAT team) performed the observations and 
reduced the \CII\ data, C.B. developed the (self)-PCA method. 
N.S., L.B., S.B. led the data interpretation and write-up, V.O., S.K., A.G.G.M.T.and T.Cs. contributed in 
discussions. S.K. performed the FUV field calculations, the HISA study and provided plots for the error calculation. 
N.S. and V.O. applied the data to the PDR toolbox and interpreted the results. 

\subsection{Competing interests} 
The authors declare no competing interests.

\subsection{Additional information} 
Correspondence and requests for materials should be addressed to Nicola Schneider or Lars Bonne.

\end{document}